Piotr Bazydło
Krzysztof Lasota
Adam Kozakiewicz

Research and Academic Computer Network NASK
Kolska 12, 01-045 Warsaw, Poland
piotr.bazydlo@nask.pl
krzysztof.lasota@nask.pl
adam.kozakiewicz@nask.pl

# Botnet fingerprinting method based on anomaly detection in SMTP conversations

**Abstract:** The paper presents the results obtained during research on detection of unsolicited e-mails which are sent by botnets. The distinction from most of the existing solutions is the fact that the presented approach is based on the analysis of network traffic – the sequence and syntax of SMTP commands observed during email delivery process. The paper presents several improvements for detection of unsolicited email sources from different botnets (fingerprinting), which can be used during network forensic investigation.

## 1. Introduction

The system of electronic mail is essential for business communication and most of the organizations cannot imagine working or even existing without e-mail accounts. On the other hand, it is also used for malicious activities. Just sending the advertising emails, besides being inconvenient and wasting various types of resources, e.g. available bandwidth (more than half of the email messages are classified as spam [1]) or disk space, carries a small risk of a security violation. Spam is commonly used during phishing campaigns and can include links to malicious websites or malicious attachments.

Classical approach to e-mail forensic investigation [2] assumes analysis of electronic message headers and contents to identify the sender. The direct source of spam usually turns out to be misconfigured/compromised mail servers or specially prepared brokers (proxies), known as open relays which allow uncontrolled sending of electronic messages. Such sources of threats can be trivially filtered using blacklists.

This in turn resulted in the development of advanced spambots - malicious programs specialized in sending spam using their own implementation or existing libraries containing SMTP (Simple Mail Transport Protocol - standard protocol used to send e-mail) client functionality. However, the modular structure of modern bots allows the same infected machines to perform other malicious activities (DDoS, bitcoin mining etc.) [3, 4].

Figure 1 illustrates the two main modes of spambot operation. In the first mode (Fig. 1A), an infected computer sends spam via the aforementioned brokers. Thereby, it is restricting the number of locations from which unwanted messages reach the target e-mail servers. In the second case (Fig. 1B), where bots are able to use specific conversation with mail servers, the number of spam sources dramatically increases, up to the number of compromised machines controlled by the botmaster. Mail servers receive messages from a huge number of sources. Filtration is no longer trivial, source e-mail address, domain or IP address of the sender are not enough.

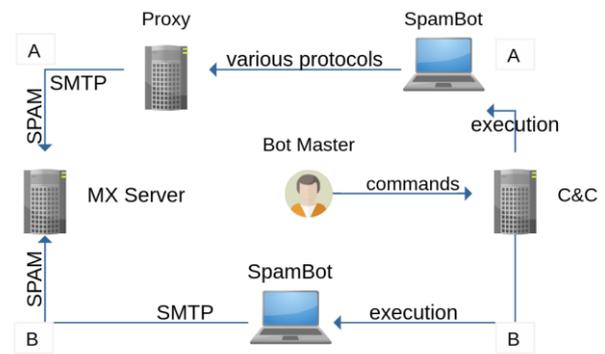

*Fig. 1. Spambot email delivery with brokers (A) and direct (B)*

A large number of scientific papers describing methods of spam detection have been published [5, 6]. These are mainly based on the analysis of the received message - its headers, content and attachments. For this purpose statistical methods [7], graph theory [8] and even machine learning algorithms [9], are used. Noteworthy are the papers of Stringhini et al. ([10] and its follow-up [11]), which present quite a different approach. Authors focus on the analysis of network communication observed during the sending of a message by a spambot, while creating a profile of so-called SMTP dialects and comparing them with normal traffic generated by benign mail sending applications. Their results show mistakes in the implementation of the SMTP protocol made by the malware authors and therefore possibility of its classification. Detection of anomalies in SMTP traffic may find application in the network forensic investigation on malicious activity of spambots [12]. Paper [13] presents in-depth analysis of spam detection through network characteristics, like IP TTL derived from TCP SYN packet or completion time of TCP three way handshake.

In this paper, we focus on the development of the algorithm of detection and classification of messages sent by spambots. The article presents additional features (section 3) and results (both spam detection and botnet fingerprinting) that we have obtained during our work on the analysis of unsolicited emails (section 4). The main part of the article is preceded by a short description of the structure of an SMTP conversation and a definition of an SMTP dialect (section 2). The paper finishes with a brief conclusion of our findings (section 5).

## 2. SMTP Dialects

In order to understand the idea of SMTP dialects analysis, some basics have to be established.

SMTP is a global standard for email transfer. This text-based protocol was defined in 1982 and updated in 2008 by RFC 5321. In simple words, it can be described as the rules of conversation between MUA (Mail User Agent, e.g. Microsoft Outlook, Mozilla Thunderbird) clients and MTA/MDA servers (Mail Transfer Agent, Mail Delivery Agent, e.g. Postfix or Microsoft Exchange Server), which should result in electronic mail delivery.

Generalized mail delivery is presented in fig. 2 and the procedure is performed in a following way:

- 1) user provides an input (message and recipients) to the MUA which transfers mail to the mail server using the SMTP protocol,
- 2a) if server is used as a relay, the MTA forwards the message to the appropriate mail server using SMTP,
- 2b) if the current MTA is the final destination of the message, mail delivery to the local mailbox is handled by the server's MDA (Mail Delivery Agent),
- 3) mail is downloaded from the proper mailbox by the MUA using POP (Post Office Protocol) or IMAP (Internet Message Access Protocol) protocols.

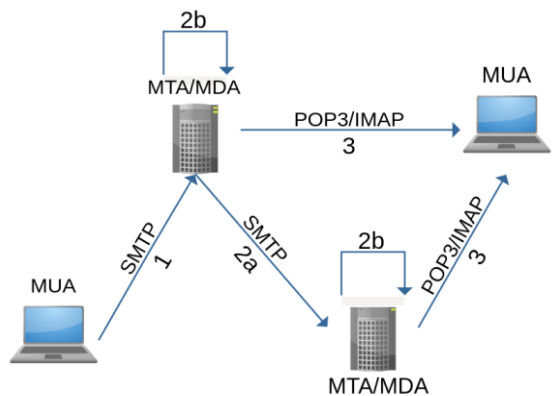

*Fig. 2. Direct and indirect email delivery.*

SMTP conversation is carried out according to the client-server model and looks the same, whether message delivery is performed between the MUA and the MTA or two different MTA's. The following listing shows a typical SMTP conversation, where an email is sent to one recipient.

```
S: 220 smtp.server.com
C: EHLO my.example.com
S: 250 smtp.server.com
C: MAIL FROM:<sender@example.com>
S: 250 2.1.0 Ok
C: RCPT TO:<recipient@server.com>
S: 250 2.1.5 Ok
C: DATA
S: 354
C: Test message.
C: .
S: 250 2.0.0 Ok
C: QUIT
S: 221 2.0.0 Bye
```

When TCP connection is successfully established, the server greets the client (with *"220"* command).

The client responds using the "EHLO" command (or "HELO" in a deprecated version of SMTP). Then, the sender of the email is provided using "MAIL FROM" command. If the sender address is accepted by the server, the recipient address is provided using "RCPT TO". Finally, the client signalizes beginning of the message through the "DATA" command and ends it using a special terminator (dot). The transmitted message should conform to the IMF (Internet Message Format) standard. Conversation ends with the "QUIT" command and the server response.

In this simple scenario, only five from eleven primary SMTP client commands are used. Every SMTP command must end with <CR><LF> (carriage return + line feed) line terminator.

Extended SMTP allows the client to use available SMTP extensions, listed by the server after successful "EHLO". For example, the client may receive information whether encrypted connections initialized by "STARTTLS" command are supported.

Like in a spoken language, there are rules that have to be obeyed in order for an SMTP conversation to be understood by both sides, but some minor grammatical differences or errors have no real impact on the conversation outcome.

If for example, the MTA recognizes both *"MAIL FROM:<me@example.com>"* and *"mail from: me@example.com"* commands properly and can distinguish the address correctly. SMTP commands are case-insensitive, and some characters like "<>" or some whitespaces are optional. Various equivalent forms of SMTP commands lead to the possibility of distinguishing different SMTP implementations and their sources.

As mentioned in the introduction section, first work towards analysis of different SMTP protocol implementations was presented by G. Stringhini et. al. [10]. Authors provided the solution called "B@bel", in which SMTP dialects derived from incoming TCP sessions are analyzed and compared with the known (benign) SMTP dialects. In later research [12], G. Stringhini et. al. used primary IMF fields like message subject or sender address to connect SMTP dialects with spam campaigns.

## 3. SMTP Dialect Extension

As our solution is based on the concept presented in [10, 11], we propose the following improvements to the already developed SMTP dialects fingerprinting:

- analysis of full SMTP conversation ("B@bel" system was limited to the investigation until *"DATA"* command),
- analysis of SMTP extensions syntax,
- analysis of IMF fields strictly linked with dialects, i.e.: "User-Agent" containing information about MUA or "Received" containing information about email travel path,
- dialect classification in three categories: benign, suspicious and malicious.

Next chapters describe the aforementioned enhancements.

### 3.1. Parsing Full Conversation and Extensions

While previous solutions assumed that the *"DATA"* expression is the last final state in every dialect and it is not necessary to parse further commands, we have decided to extend analysis to full SMTP conversation. This can be relevant, as there may be inconsistencies in e.g. "QUIT" commands. One of the most surprising behaviors is a special usage of the *"RSET"* command after sending an email to send more emails during the same TCP session (even 10 000 emails in a single TCP connection were observed). In addition, we have provided a possibility to check the syntax of SMTP extensions described in RFC 5321.

In summary, we have developed three different operation modes for our SMTP dialects analyzer:

- method 0 (M0): parse conversation until the "DATA" command, like "B@bel" for reference purpose,
- method 1 (M1): parse the whole conversation without SMTP extensions,
- method 2 (M2): parse the whole conversation with SMTP extensions.

### 3.2. IMF Extension

As mentioned before, IMF is a standard syntax for electronic mail messages. Its header consists of different fields e.g.: "From", "To", "Subject" or "CC". We believe that several IMF fields may be successfully used in dialect analysis and botnet fingerprinting.

The first field included in analysis is responsible for informing the MTA about the MUA client type and version. Although this field is optional, it is a good practice to provide one. There are several names for this field in common use, though the most popular are probably "User-Agent" and "X-Mailer". Our classifier compares the actual SMTP dialect with the client specified in IMF. For example, we have obtained email messages whose dialects match a legitimate MUA – Mozilla Thunderbird. However, the analysis of the IMF header showed that an "User-Agent" field was not even specified or specified like in another MUA e.g. Outlook 2010. On this basis, we can treat these messages as suspicious because of the inconsistency in dialect. The same reasoning can be applied to MTA dialects.

Another information used in our research included the IMF "Received" trace field. It is supposed to inform the current MTA about the real route of an email. If the email is received by an MTA for delivery to another server, it has to insert a trace in the form of a "Received" field. In case of direct email delivery, there should not be any "Received" trace in IMF. According to this, we are able to determine whether the message is originating from an MTA or a MUA.

For example, a message with a dialect matching only a legitimate MUA (direct delivery), may state in the IMF that the email was forwarded by an MTA. This can be treated as an attempt of spoofing (client pretending to be a server) and should mark the source of message as suspicious. This relation can be reversed (and indeed such attempts were observed), with the server claiming to be a client.

### 3.3. Classifier

Our prototype solution was implemented in Python 2.7 and based on a prepared knowledge database. We have classified senders of emails in three categories:

- benign – known MUA and MTA; during selection of clients, we have taken into account the email client market statistics [14]. According to the data from January $1^{st}$ - February $1^{st}$ 2017 our selection of clients covers more than 90% of current email traffic,
- suspicious – known SMTP client implementations on different platforms in different programming languages,
- malicious – malicious dialects are collected from three main sources: real bots traffic samples provided by CERT Polska, real bots traffic from "Malware Capture Facility Project" and manual analysis of traffic received by a

| MUA | |
| --- | --- |
| Benign | Suspicious |
| Mozilla Thunderbird, Microsoft Outlook, Clawmails, Evolution, Windows LiveMail, Opera Mail, The Bat!, Pegasus Mail, Apple Mail, iPhone/iPad Mail, PHPMailer, Google Android Mail Client | Python smtplib, C/C++ libcurl, C/C++ poco, JavaMail, C# net.mail, Perl NET::SMTP, AutoIt InetSmtpMail, Powershell Send-MailMessage, VBA SMTP library (used in Microsoft Excel) |
| Benign MTA | |
| Postfix, Windows Exchange Server, Gmail, Zimbra | |
| Bot samples | |
| Geodo/Feodo, Htbot, Kelihos, Sality, Upatre, Vawtrak, Zbot | |

real-operating spamtrap.

*Table 1. Analyzed benign and suspicious MUAs and MTAs together with bot samples*

In our solution, we used SMTP command patterns and regular expressions in order to retrieve general states of SMTP dialects (Table 2.). Our regular expressions include detection of parameters like domains, email addresses, IP addresses and many others. If a word included in the SMTP command is recognized by neither regular expressions nor command patterns, it is written in our syntax as *"text"*. Unlike our predecessors, we do not create state machines of dialects. Instead, we are obtaining hashes of an ordered list of obtained states. Classification itself is based on hashes comparison and rules checking (IMF extensions).

| | Received command | Obtained state |
|---|---|---|
| S | 220 hostname | 220 |
| C | EHLO [127.0.0.1] | EHLO space [ IPv4 ] <CR> <LF> |
| S | 250 Ok | 250 |
| C | Mail FROM:<send@mail.pl> | Mail space FROM : < email > <CR> <LF> |
| S | 250 2.1.0 Ok | 250 2.1.0 |
| C | Rcpt To: <recipient> | Rcpt space To : space < text > <CR> <LF> |
| S | 250 2.1.5 Ok | 250 2.1.5 |
| C | DATA | DATA <CR> <LF> |
| S | 354 Ok | 354 |
| C | My test message. | message |
| S | 250 2.0.0 Ok | 250 2.0.0 |
| C | quit | quit <CR> <LF> |
| S | 221 2.0.0 Bye | 221 2.0.0 |

*Table 2. Exemplary conversion of SMTP commands into dialect states.*

## 4. The Results

All presented information was obtained using our testing infrastructure, which is presented in Fig. 3. The spamtrap is responsible for capturing spam from various sources. While we capture incoming network traffic on a network device, the spamtrap derives session keys (pre-master secrets) from encrypted TCP streams. Tshark software is used for data decryption and separation of different TCP streams. Parsing and classification software based on the data collected in the knowledge base saves its results in the storage. During system operation the administrator can decide whether to add one of the unknown dialects to the knowledge database (this operation can be performed later, after further analysis).

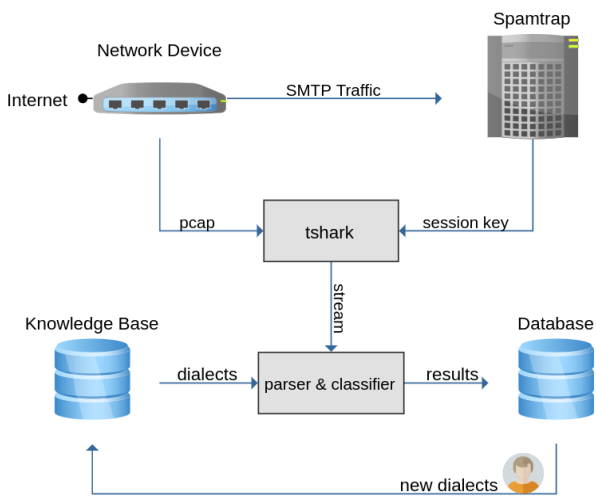

*Fig. 3. Testing Environment*

### 4.1. Knowledge database

The knowledge database is composed of dialects derived from clients and real bot samples included in Table 1 and dialects coming from network traffic incoming to our spamtrap. Table 3 presents the number of benign, suspicious and malicious dialects distinguished by different operation modes.

| Operation mode | M0 | M1 | M2 |
|---|---|---|---|
| Known benign | 29 | 50 | 70 |
| Known suspicious | 4 | 19 | 23 |
| Known malicious | 34 | 44 | 46 |
| Total | 67 | 113 | 139 |

*Table 3. Number of distinguished dialects for every mode*

In accordance with the assumption, the more complex our analysis is, the more dialects we can distinguish. M2 recognizes more than twice as many dialects as M0. Many differences in benign or suspicious clients occur after the *"DATA"* command, as the big difference between the number of benign dialects for M0 and M1/M2 shows.

In every operation mode, the number of command variants significantly increases the number of known dialects. It indicates that there are a lot of mistakes / inaccuracies in botnets' SMTP implementations. However, the vast majority of malicious classifications comes from the first part of SMTP conversation, e.g. the "MAIL FROM" command has four benign variants, whereas up to twelve malicious variants were distinguished, depending on the method. Additionally, we have discovered several unknown client commands, which are improper according to the SMTP standard, e.g. *"220 domain ESMTP Welcome"* – an attempt of a client at sending server commands.

### 4.2. Result Statistic

Our data set consists of 32470 conversations, from which 13017 are scans on port 25 and others are conversations finished with a successful email transfer. It should be noted that every single obtained conversation in reality originated from a malicious source. As the spamtrap is not used for regular email transfer, the only messages incoming are SPAM. According to this, we can obtain only two classifications: TP (True Positive) when spam is detected correctly and FN (False Negative) if not. However, spam FP (False Positive) and TN (True Negative) rates are not really relevant from the botnet fingerprinting perspective, as we are focused only on dialects and IMF matching.

This chapter presents simple numerical results for all streams from the validation set. We have divided the obtained results into three different categories (table 4):

- "UNKNOWN" denotes conversations in which every SMTP command is known from legitimate MUAs and MTAs, although the command sequence is unknown (not correctly matched with any dialects from the database), thus we are not able to categorize it,
- "MALICIOUS" denotes malicious conversations which were collected from manual analysis of spambots' traffic, traffic classified as originating from malicious activities and real bot samples,
- "KNOWN" denotes conversations which are correctly matched with the database of legitimate dialects.

Additionally, we have used four metrics:

- Number of samples – the number of conversations matching the current category,
- Ratio of total - ratio of matched samples and all captured SMTP conversations,
- IMF inconsistency – number of alerts triggered for conversations matching current category,
- Ratio of alerts – ratio of alerts triggered and number of all categorized conversations.

Alerts are triggered if there is an inconsistency between the dialect and the IMF header fields, described in chapter 3.2 (spoofing MUA client type or incorrectly matched email trace fields with MUA/MTA). Samples which triggered an alert should be treated as malicious.

In table 4, three different operation modes described in chapter 3.1 are compared. As almost 62% of incoming SMTP traffic does not match any legitimate dialect, it can be immediately treated as originating from malicious sources while using our M1 and M2 operations mode (table 4 classification part). On the other hand, M0 based on the solution [10] is able to classify about 59% of traffic as malicious.

If we treat inconsistency between the dialects and the IMF headers as at least suspicious, it turns out that the number of False Negatives can be significantly reduced. Analysis of the results shows that our proposed IMF based extension has greatly improved classification results with a 18.6%, 33,4% and 19.5% True Positive enhancement for M1, M2 and M0 respectively. M2 connected with IMF analysis reached an impressive 95 % TP rate. In case of M1 and M0 known dialects, all alerts were trigged by incorrectly specified "X-Mailer" or "User-Agent" IMF fields. In case of M2, 4832 alerts were triggered by the wrong mailer fields, while 1675 alerts were triggered by client/server spoofing (analysis of "Received" IMF field).

| RESULTS | | | |
|---|---|---|---|
| Type | UNKNOWN M1 | UNKNOWN M2 | UNKNOWN M0 |
| Number of samples | 95 | 506 | 505 |
| Ratio of total (%) | 0,5 | 2,6 | 2,6 |
| IMF inconsistency | 87 | 88 | 87 |
| Ratio of alerts (%) | 91,6 | 17,4 | 17,2 |
| Type | MALICIOUS M1 | MALICIOUS M2 | MALICIOUS M0 |
| Number of samples | 11459 | 11465 | 11474 |
| Ratio of total (%) | 58,9 | 58,9 | 59,0 |
| IMF inconsistency | 6711 | 6715 | 6716 |
| Ratio of alerts (%) | 58,6 | 58,6 | 58,5 |
| Type | KNOWN M1 | KNOWN M2 | KNOWN M0 |
| Number of samples | 7899 | 7482 | 7474 |
| Ratio of total (%) | 40,6 | 38,5 | 38,4 |
| IMF inconsistency | 3792 | 3608 | 6507 |
| Ratio of alerts (%) | 48,0 | 48,2 | 87,1 |
| CLASSIFICATION | | | |
| Mode | M1 | M2 | M0 |
| TP (%) | 59,4 | 61,5 | 61,6 |
| FN (%) | 40,6 | 38,5 | 38,4 |
| Mode | M1 with IMF ext. | M2 with IMF ext. | M0 with IMF ext. |
| TP (%) | 78,9 | 80,1 | 95,0 |
| FN (%) | 21,1 | 19,9 | 5,0 |

*Table 4. Comparison of results in different operation modes*

Solution based on SMTP dialects analysis clearly provides good results. However, there are many legitimate MUAs and it is almost impossible to collect every benign dialect. That can lead to the improper classification of a legitimate message as spam (False Positive). This is not a concern when analysis is focused exclusively on botnet fingerprinting. In order to avoid a growing number of False Positives, the database should be constantly updated. Nonetheless, we have provided the IMF extension which is able to detect attempts of spoofing. About 60% of obtained conversations with unknown / malicious dialects were pretending to be a different MUA than they were in reality (mostly pretending to be an Apple device or Microsoft Outlook). Our database of legitimate dialects included clients, which according to the statistics correspond to more than 90% of email traffic (on the basis of 1.25 billion messages) [14].

### 4.3 Botnet fingerprinting

Botnet fingerprinting can be performed by matching of bot dialects and IMF fields with dialects derived from network traffic incoming to the SMTP server. In around 19500 successful SMTP conversations, following botnets were fingerprinted in M1 mode:

- Vawtrak: 3872 messages from 2 IP addr.,
- Kelihos: 2816 messages from 627 IP addr.,
- Htbot: 879 messages from 669 IP addr.,
- Zbot: 1 message from 1 IP addr.

Although we have limited access to real bot samples, almost 40% of traffic was successfully assigned to botnets. Moreover, TCP/IP headers analysis [13] can be additionally performed, in order to increase accuracy of fingerprinting, as implementations of SMTP in some bots may be similar to the legit SMTP dialects.

### 4.4 Interesting cases

Some selected cases of unusual mistakes in SMTP implementations will now be presented.

First example concerns the bot responsible for a malicious eFax spam campaign [15]. In the IMF header, the bot sample signalized usage of four MUAs (depending on the message): Apple Mail, iPhone Mail, iPad Mail, iPod Mail. However, the bot SMTP dialect differs from real dialects derived from the Apple devices:

- the malicious dialect greets with *"HELO domain"*, the benign dialect greets with *"EHLO [IP]"*,
- only the malicious dialect inserts space character between *":"* and *"<"* characters in *"MAIL FROM"* and *"RCPT TO"* commands,
- only the benign dialect uses the *"QUIT"* command.

Second example presents a big, although hard to notice mistake. In this dialect, line terminators are $<LF>$ instead of $<CR><LF>$. Such an implementation is incompatible with SMTP standard RFC 5321, and can lead to problems in communication between different MUAs and MTAs. However, the Postfix MTA can process commands with *line feed* terminators, so spam can be propagated without hindrance.

Next example describes an almost successful implementation of a benign-looking SMTP dialect with just one seemingly negligible error. The malicious client responsible for spam propagation finishes the conversation using a command parsed as "QUIT space

<CR><LF>" – inserting a space character between the command and line terminator characters. This is an unusual behavior, not present in any analyzed benign dialect. Moreover, no other commands include this space character. It is probably an outcome of a typo, thus allows to doubt in authenticity of email message source.

Additionally, using the described methodology of SMTP dialect analysis, it is possible to fingerprint bots/systems responsible for scanning SMTP ports. Moreover, we observed that bots are checking if particular security mechanisms are handled by the SMTP server, like starting TLS connection with "STARTTLS" command or user authentication.

## 5. Conclusion

The paper discusses an important section of digital forensics – botnet fingerprinting. Our work is an overview and considerable extension of the method proposed in [10], based on an analysis of the SMTP implementations in clients / bots. Our extensions and methods were validated on the network traffic incoming to the real-operating spamtrap. In addition, we had access to the real bot samples.

We have proven that it is beneficial to extend the network traffic analysis to the full SMTP conversation. Two additional analyses based on proper interpretation of chosen IMF fields have led to a significant decrease of False Negative classifications (from around 40% to even 5%). Additionally, the IMF header analysis can also decrease the number of skipped samples (unknown dialects) and classify them as malicious when an attempt of spoofing is detected.

Having access to 7 SMTP dialects derived from real bots, our method was able to fingerprint 4 different botnets in 19500 spam messages. Three other bot dialects were not spotted in the spamtrap traffic.

Spam detection methods are usually paired with each other and classification is rarely performed on the basis of a single solution. It seems especially attractive to connect our method with the others based on the analysis of the message contents, as it will provide the symbiosis of effective spam detection methods. However, our future research will be focused on matching malicious dialects and sources with particular spam campaigns. This may be important in fingerprinting not only single bots, but the whole botnets.

In summary, we have proposed several improvements to the idea of botnet fingerprinting and spam detection achieved by analysis of SMTP implementation. Our results show high accuracy of the aforementioned methods and presented extensions. Initial tests and research proved that our solution can be successfully used in both botnet fingerprinting and spam detection.

## Acknowledgment

The research leading to these results has received funding from the European Union Horizon 2020 Programme (H2020-DS-2015-1) under grant agreement n° 700176 (project SISSDEN).

## Bibliography

[1] D. Gudkova, M. Vergelis, N. Demidova, T. Shcherbakova, Spam and phishing in Q2 2016, Kaspersky Lab, August 18, 2016.

[2] Banday, M. Tariq. "Techniques and Tools for Forensic Investigation of E-mail." International Journal of Network Security & Its Applications 3.6 (2011): 227.

[3] Karim, A., Salleh, R.B., Shiraz, M., Shah, S.A.A., Awan, I., Anuar, N.B., Botnet detection techniques: review, future trends, and issues, Journal of Zhejiang University: Science C, Volume 15, Issue 11, 2014, pp. 943-983.

[4] John J. P., Moshchuk A., Gribble S. D., Krishnamurthy A., Studying Spamming Botnets Using Botlab, USENIX Symposium on Networked Systems Design and Implementation NSDI, 2009.

[5] Spirin, Nikita, and Jiawei Han. "Survey on web spam detection: principles and algorithms." ACM SIGKDD Explorations Newsletter 13.2 (2012): 50-64.

[6] Khan, W.Z., Khan, M.K., Bin Muhaya, F.T., Aalsalem, M.Y., Chao, H.-C., A Comprehensive Study of Email Spam Botnet Detection, IEEE Communications Surveys and Tutorials, Volume 17, Issue 4, Fourth Quarter 2015, pp. 2271-2295.

[7] Meyer T. A., Whateley B., SpamBayes: Effective open-source, Bayesian based, email classification system, Collaboration, Electronic messaging, Anti-Abuse and Spam Conference CEAS, 2004.

[8] P. Haider and T. Scheffer, "Finding botnets using minimal graph clusterings," in Proc. 29th Int. Conf. Mach. Learn., Edinburgh, U.K., Jun. 26–Jul. 1, 2012, pp. 847–854.

[9] Vural I., Venter H.S, Using Network Forensics and Artificial Intelligence Techniques to Detect Bot-nets on an Organizational Network, Seventh International Conference on Information Technology: New Generations (ITNG), 2010, pp. 725-731.

[10] Stringhini, G., Egele, M., Zarras, A., Holz, T., Kruegel, C., & Vigna, G. (2012, August). B@bel: Leveraging Email Delivery for Spam Mitigation. In USENIX Security Symposium pp. 16-32.

[11] Stringhini, G., Hohlfeldy, O., Kruegel, C., Vigna, G., The harvester, the botmaster, and the spammer: On the relations between the different actors in the spam landscape, 9th ACM Symposium on Information, Computer and Communications Security, ASIA CCS 2014; Kyoto; Japan; 4 June 2014, pp. 353-363.

[12] Joshi, R. C., and Emmanuel S. Pilli. Fundamentals of Network Forensics: A Research Perspective. Springer, 2016.

[13] Tu, O., Ray, S., Allman, M., Rabinovich, M., "A large-scale empirical analysis of email spam detection through network characteristics in a stand-alone enterprise." Computer Networks 59 (2014), pp. 101-121

[14] Litmus labs, Email Client Market Share statistics https://emailclientmarketshare.com/

[15] Malware-traffic-analysis blog, http://malware-traffic-analysis.net/2017/02/07/index.html